\documentstyle[nato,numreferences,epsf,graphicx,times,mathptm]{crckapb}
\newcommand{\beq}{\begin{equation}}
\newcommand{\eeq}{\end{equation}}
\newcommand{\bea}{\begin{eqnarray}}
\newcommand{\eea}{\end{eqnarray}}
\newcommand{\ket}[1]{|#1\rangle}
\newcommand{\bra}[1]{\langle#1|}

\newcommand{\E}{{\cal E}}

\newcommand{\vx}{\vec{x}}

\renewcommand{\ni}{\noindent}

\renewcommand{\d}{\delta}

\renewcommand{\b}{\beta}
\renewcommand{\a}{\alpha}

\newcommand{\m}{\mu}

\newcommand{\s}{\sigma}

\newcommand{\oh}{\frac{1}{2}}

\newcommand{\dg}{\dagger}
\newcommand{\non}{\nonumber}

\newcommand{\rf}[1]{(\ref{#1})}
\newcommand{\ra}{\rightarrow}

\newcommand{\pa}{\partial}
\begin{opening}
\title{The Gluon Chain Model Revisited}
\author{Jeff Greensite}
\institute{Physics and Astronomy Dept., San Francisco State University,\\
           San Francisco, CA 94117, USA, and \\
           Theory Group, Lawrence Berkeley National Laboratory,\\
           Berkeley, CA 94720, USA}
\runningtitle{Gluon Chain Model}
\runningauthor{J.\ GREENSITE}
\end{opening}
\begin{document}
%
%
\begin{abstract}
   I describe how the gluon chain model of QCD string formation meets
a number of criteria which are required of any theory of the confining
force, including: the correct center dependence and (at large-N)
Casimir scaling of the string tension, the logarithmic broadening of
the QCD flux tube, and the existence of a L\"uscher term in the static
quark potential.
\end{abstract}
%
%
\renewcommand{\thefootnote}{\fnsymbol{footnote}}
\footnotetext[1]{
Talk presented at the Nato Advanced Research Workshop on {\sl Confinement,
Topology, and Other Non-Perturbative Aspects of QCD}, Star\'a Lesn\'a,
Slovakia, January 21-27, 2002.
Supported in part by
the US Department of Energy, Grant No.\ DE-FG03-92ER40711.}
\renewcommand{\thefootnote}{\arabic{footnote}}

\section{Introduction}

   The gluon chain model [1-4] is a picture of the formation and
composition of the QCD string in terms of the perturbative excitations
of the theory.  In this talk I would like to discuss some recent work
\cite{chain} in this area, which was carried out in collaboration with
Charles Thorn.

   Normally gluon (particle) excitations are useful for describing
high-energy scattering processes, while non-perturbative effects, such
as confinement and chiral symmetry breaking, are usually ascribed to
some special class of field configurations with particular topological
properties.  However, it is not excluded that particle and field
descriptions of various phenomena in QCD can overlap in some cases.
An example is the color screening of higher representation Wilson
loops.  In this case one is easily convinced, just by thinking about
string breaking due to gluon pair production, that the asymptotic
string tension of a higher-representation loop can depend only on the
transformation properties of the representation with respect to the
center subgroup (i.e.\ on the ``N-ality'' of the representation).  For
an adjoint string in particular, one ends up with two gluelump states,
each consisting of a gluon bound to a heavy source.  On the other
hand, from the ``field'' point of view, the area law falloff of the
Wilson loop is due to large scale vacuum fluctuations of some kind.
The N-ality dependence of the string tension tells us that these
large-scale field fluctuations must induce fluctuations in Wilson loop
holonomies only among the center elements, rather than coset elements,
of the SU(N) gauge group.  This simple fact implies (very strongly, in
my opinion) the existence of a center vortex mechanism of some kind.

\begin{figure}[h!]
\centerline{\scalebox{0.40}{\includegraphics{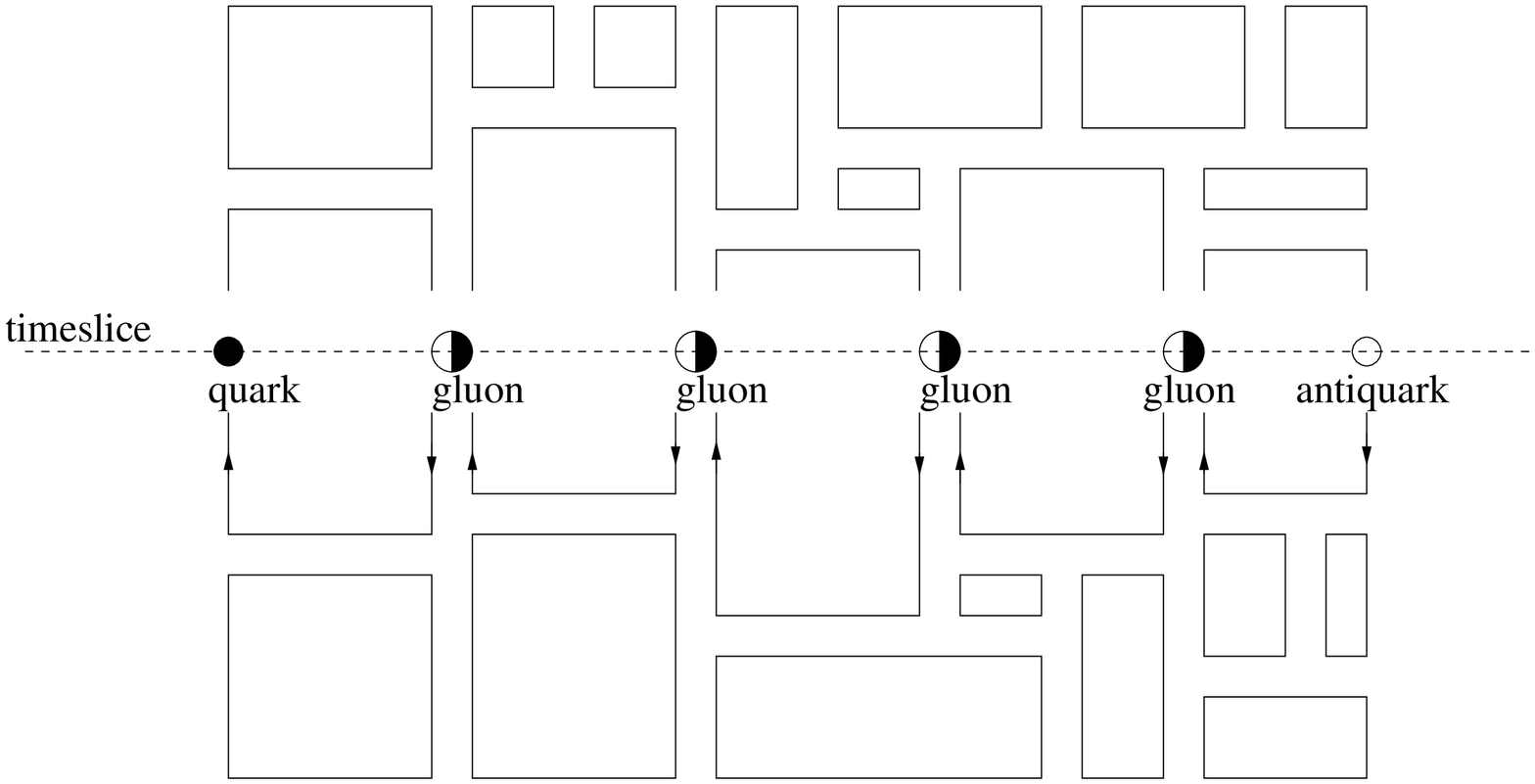}}}
\caption{The gluon chain as a time slice of a planar diagram (shown here
in double-line notation).
A solid (open) hemisphere indicates a quark (antiquark) color index.}

\label{planar}
\end{figure}

    In principle we then have two complementary ways of understanding
the N-ality dependence of the QCD string tension: (i) in terms of
color-screening by constituent gluons; and (ii) in terms of the vortex
confinement mechanism.  But if we can picture string \emph{breaking}
in terms of particle excitations, is it also possible to describe
string \emph{formation} in terms of constituent gluons?  As it
happens, large-$N_c$ considerations lead to just such a description.
Consider a very high-order planar Feynman diagram contributing to
a Wilson loop expectation value, as shown in Fig.\ \ref{planar}. A time-slice
of the diagram reveals a sequence of gluons, interacting only with
their nearest neighbors in the diagram.  If we take this picture
seriously, it suggests that the QCD string might be regarded, in some
gauge, as a ``chain'' of constituent gluons, with each gluon held in
place by its attraction to its two nearest neighbors in the chain.

\begin{figure}[h!]
\centerline{\scalebox{0.40}{\includegraphics{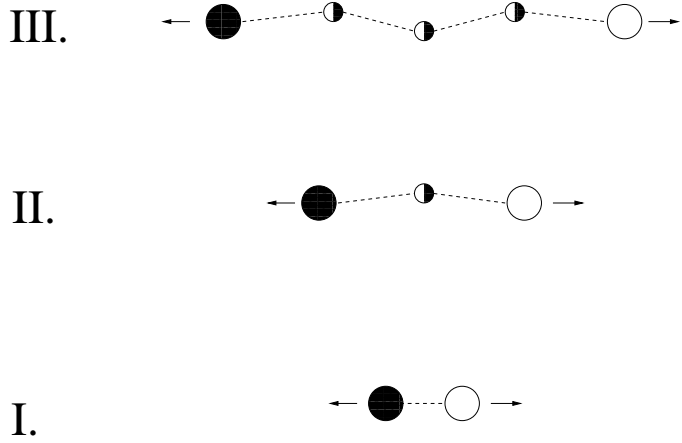}}}
\caption{Constituent gluons (small circles) appear as quarks separate, to keep
the average color charge separation below some maximum value.
Dotted lines indicate nearest-neighbor interactions.}
\label{single}
\end{figure}

   The linear potential in this ``gluon chain'' model comes about
in the following way:  As the heavy quarks separate, we expect that at
some point the interaction energy increases rapidly due to the running
coupling.  Eventually, it becomes energetically favorable to insert
a gluon between the quarks, to reduce the color charge separation.
As the quarks continue to separate, the process repeats, as shown in
Fig.\ \ref{single}, and we end up with a chain of gluons.  The average 
gluon separation $R$ along the axis joining the quarks is fixed, 
regardless of the quark separation $L$,
and the total energy of the chain is the energy per gluon
times the number $N$ of gluons in the chain, i.e.
\beq
      E_{chain} \approx N E_{gluon}
                = {E_{gluon} \over R} L
                = \s L
\eeq
where $E_{gluon}$ is the (kinetic+interaction) energy per gluon,
and $\s = E_{gluon}/R$ is the string tension.  In this picture,
the linear growth in the number of gluons in the chain is the origin 
of the linear potential.
   
   Every theory of confinement aims at explaining the linearity of
the static quark potential at large distances.  However, it is by now 
well understood that linearity
is only one of several conditions that a theory of the confining force
must satisfy.  The requirements also include:
\begin{enumerate}
\item {\bf Casimir Scaling}.  At intermediate distance scales, the
string tension should be proportional to the quadratic Casimir of
the quark color group representation.
\item {\bf Center Dependence}. Asymptotically, the string tension
should depend only on the N-ality, not the Casimir, of the
quark color group representation.
\item {\bf String Behavior:}
  \begin{itemize}
  \item {\bf Roughening}.  There is a logarithmic broadening of the flux
  tube with quark separation $L$.
  \item {\bf L\"uscher term}.  There is a universal $-c/L$ term in the
  static quark potential at large distances.
  \end{itemize}
\end{enumerate}

  Taken together, this is a challenging set of conditions.  
The abelian projection theory, for example, has difficulties with
both Casimir scaling \cite{Lat96} and center dependence \cite{j3}.  
Other proposals, which seek to derive confinement from a
$1/k^4$ behavior of the gluon propagator in some gauge, will
run into trouble when confronted with roughening and the 
L\"uscher term.  So does the gluon chain model, which is yet another
idea about confinement, do any better in meeting these conditions?

\begin{figure}[h!]
\centerline{\scalebox{0.40}{\includegraphics{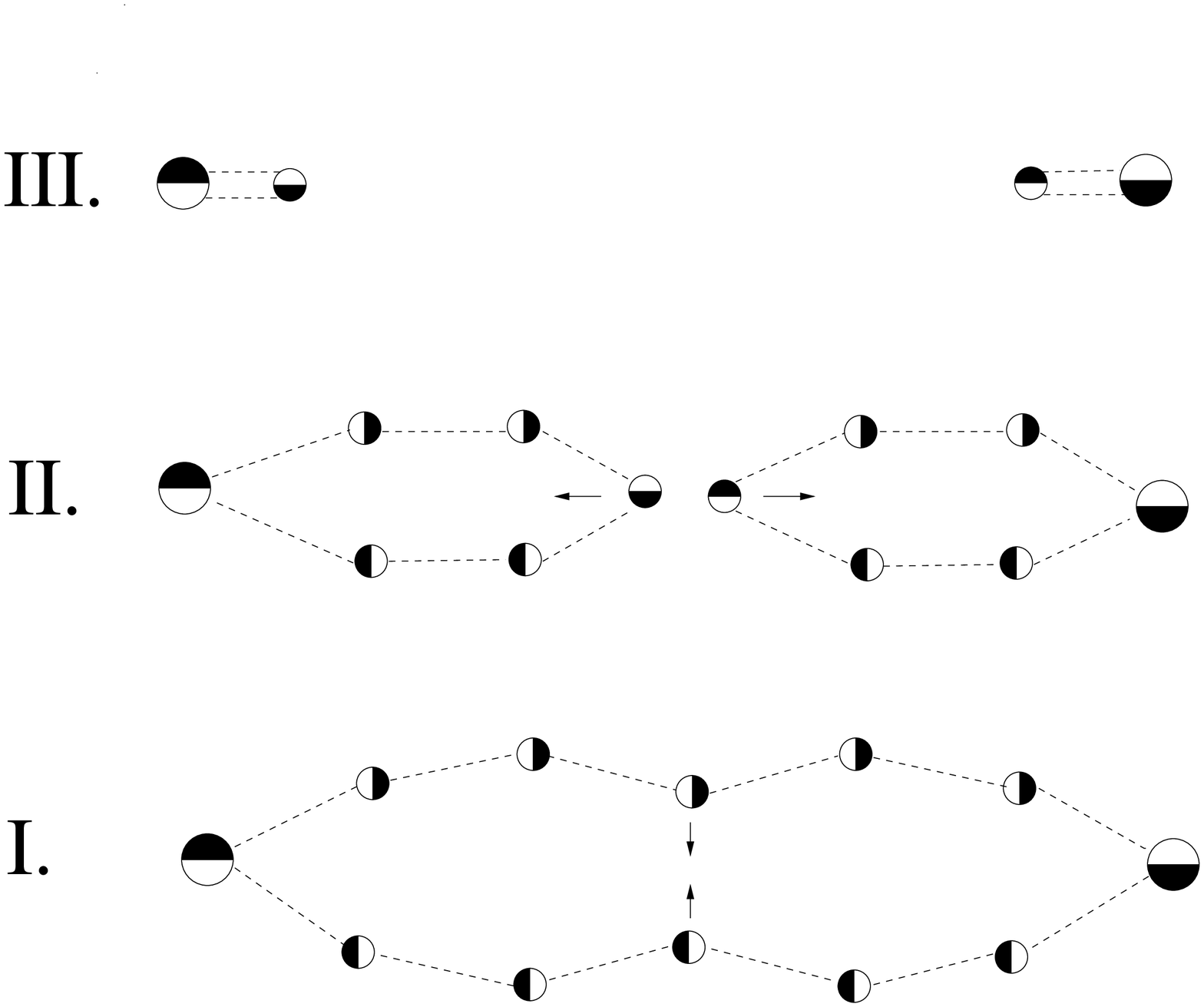}}}
\caption{Adjoint string-breaking in the gluon chain model.
Two gluons in separate chains (I) scatter by a contact interaction,
resulting in the re-arrangement of color indices indicated in II.
This corresponds to chains starting and ending on the same heavy source.
The chains then contract down to smaller ``gluelumps'' (III).}
\label{double}
\end{figure}

   Having already discussed the linearity of the static potential, 
let us consider Casimir scaling. In any SU($N_c$) representation $ r$, 
the group character can be expressed as a product
\beq
      \chi_r[g] \propto \Bigl(\chi_F[g]\Bigr)^n 
                  \Bigl( \chi^*_F[g] \Bigr)^{\overline{n}}
           ~ + ~ \mbox{sub-leading terms} .
\eeq 
where $F$ denotes the fundamental representation.
By factorization at large-$ N_c$, a Wilson loop in 
representation $ r$ has
a string tension
\beq
     \s_r = M_r \s_F
\eeq
at $ N_c \ra \infty$, where $ M_r = n + \overline{n}$.  
In this limit,
the quadratic Casimir is $ C_r = M_r N_c/2$.  Exact Casimir scaling is 
therefore a property of the planar limit.  The gluon chain model,
which is motivated by large-$N_c$ considerations, inherits this property.
In the gluon-chain model, there are $ M_r$ chains terminating
at each heavy source, and these are non-interacting at 
$ N_c \ra \infty$.  The total energy of the system is then proportional
to the number of chains times the length of each chain, and in 
this way Casimir scaling
\beq
       \s_r \propto C_r ~~~~~(N_c \ra \infty)
\eeq
is obtained at large $N_c$.
The situation for heavy adjoint-representation quarks, which have two
chains between them, is shown in Fig. \ref{double}(I).

   The appropriate center dependence of the string tension
is due to color screening.  Screening 
is accomplished by a ($ 1/N_c^2$ suppressed)
contact interaction between constituent gluons in different
chains, leading to string-breaking processes such as those shown in
Fig.\ \ref{double}.

   As for string behavior, the gluon chain is clearly some sort of 
discretized string, so string-like properties are not entirely unexpected.  
I will return to this topic in section 4, below.

\section{The Force Renormalization Scheme}

\begin{figure}[h!]
\centerline{\scalebox{0.40}{\includegraphics{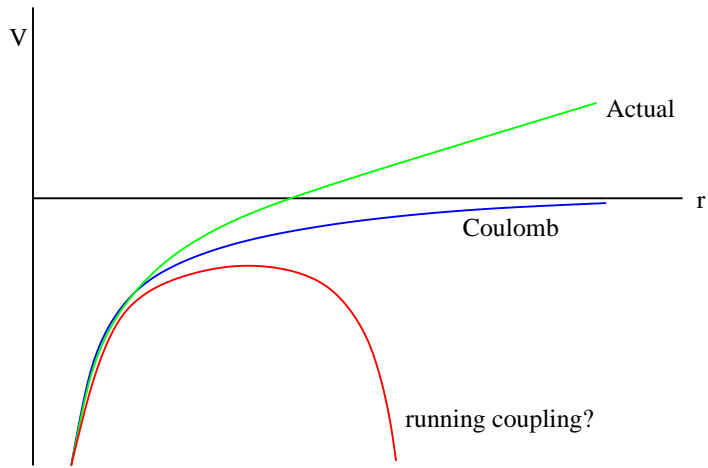}}}
\caption{A running coupling in the V-scheme which is monotonically
increasing takes the potential in the negative-V direction, away from
the actual potential.}
\label{v3}
\end{figure}

\begin{figure}[h!]
\centerline{\scalebox{0.50}{\includegraphics{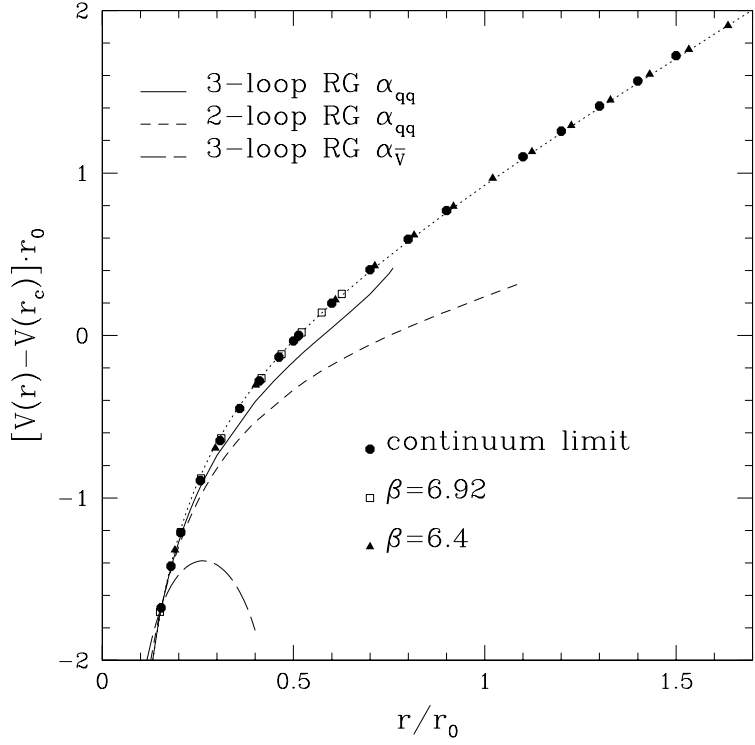}}}
\caption{The static potential in the F and V-renormalization
schemes, compared to numerical data.  This figure is taken
from Necco and Sommer, ref.\ \protect \cite{NS}.}
\label{nspot}
\end{figure}

   The growth of the running coupling with color charge separation is an
essential ingredient of gluon chain picture, but the relationship of
the running coupling to the interaction potential, which depends on the
choice of renormalization scheme, is an important consideration.  Suppose
one defines the running coupling in terms of the static potential via
\beq
V_{q\bar q}(R)=-\left(1-{1\over N_c^2}\right){N_c\alpha^V_s(R)\over 2R}.
\eeq
This is the $V$-renormalization scheme.
But if $\a_s^V$ grows monotonically, the resulting behavior of $V(R)$
is the \emph{opposite} of what
is required for confinement, behaving roughly like the curve labeled
``running coupling'' in the sketch shown in Fig.\ \rf{v3}. 
In order that the potential becomes positive at some point, as it does
in the actual potential found in numerical simulations,
it is obviously necessary that $\a_s^V(R)$ becomes smaller
and eventually \emph{changes sign} as $R$ increases.
This is asking a lot of perturbation theory, which tends in the
opposite direction.
A better renormalization scheme, advocated by Grunberg \cite{mrx},
Sommer \cite{Sommer}, and also by Thorn, is to define
the running coupling in terms of the force
\beq
|F(R)|=\left(1-{1\over N_c^2}\right){N_c\alpha_s^F(R)\over 2R^2}
\eeq
and derive the potential at intermediate distances from integration
\beq
    V_{q\bar q}(R) = V_{q\bar q}(R_A) + \int_{R_A}^R dR ~ \Bigl| F(R) \Bigr|
\eeq
In this ``F-scheme'' the running coupling doesn't have to change
sign, and in fact the three loop term in the beta function
\beq
\b(g) = -g^3 \left( {11\over (4\pi)^2}  + {102\over (4\pi)^4} g^2
        + {1.65 \over (4\pi)^3} g^4 ...\right)
\eeq
is substantially smaller (by a factor of 2.6) 
in the F-scheme than the V-scheme.  
The three-loop potential has been computed in the F-scheme
by Necco and Sommer in ref.\ \cite{NS}.  I have extracted Fig.\
\ref{nspot} below from their article; this figure displays the perturbative
potential compared to the corresponding Monte Carlo data.  The
parameter $r_0 \approx 0.5$ fm is the Sommer scale.
Notice that the F-scheme gives a remarkably good fit to the
numerical data almost up to the Landau pole, while the V-scheme results
in a very poor fit to the data in this interval.

   According to a rigorous theorem \cite{Bachas}, the static potential
must be concave downwards; i.e.\ the force cannot increase with distance.  
This implies that the the perturbative 3-loop result must certainly break 
down at quark separation $L \approx 0.6 r_0 \approx 0.3$ fm in the 
F-scheme. In the gluon-chain model, the rising force is avoided by inserting
constituent gluons between the quarks.  From the 
breakdown of perturbation theory
at $L=0.6 r_0$, we can estimate that the inter-gluon separation along
the line joining the quarks is $R=0.3 r_0$.

\section{A Variational Approach}

  Let $ \Psi_0[A]$ be the QCD vacuum
wavefunctional.  An excited state can be written as

\beq
      \Psi_{ex}[A] = Q[A] \Psi_0[A] .
\label{excited}
\eeq
e.g.\ for a glueball
\beq
      \Psi_G[A] = \sum_{N=1}^{N_{max}} \Psi^{(N)}_G[A]
\eeq
where 
\bea
   \Psi^{(N)}_G[A] & =& \left\{ \int d\vx_1 d\vx_2 ... d\vx_N 
   ~f_{\m_1 \m_2 ... \m_N}(\vx_1,\vx_2,...,\vx_N)\right.\nonumber\\
   &&\hskip1cm  \left. \phantom{\int}  \mbox{Tr}
    A_{\m_1}(\vx_1) A_{\m_2}(\vx_2) ... A_{\m_N}(\vx_N) \right\}
     \Psi_0[A] 
\eea
is the N constituent gluon state.  The energy is
\beq
   E = {\langle \Psi_{ex}|H|\Psi_{ex}\rangle 
         \over \langle \Psi_{ex} |\Psi_{ex} \rangle } - \langle H \rangle 
\eeq
Defining
\beq
        Q_t \equiv Q[A(\vx,t)]
\eeq
it is easy to show that
\beq
    E =  {-\oh} \lim_{T\ra 0} {d \over dT} \log
           \langle Q^\dg_T Q_{-T} \rangle 
\label{Eex}
\eeq
where $\langle ...\rangle$ represents the Euclidean VEV.

   The idea is to choose $Q$ as a function of a few parameters, and
then determine the optimal parameters by minimizing the value of $E$ 
computed perturbatively \cite{GH}.  For
the gluon chain, we propose to use
\bea
  Q_{chain}[A] &=& 
q^{a_1}({\bf 0}) \Bigl\{\int d\vx_1 d\vx_2 ... d\vx_N 
   ~\psi_{\m_1 \m_2 ... \m_N}(\vx_1,\vx_2,...,\vx_N) 
\non \\
    & &  A^{a_1 a_2}_{\m_1}(\vx_1) A^{a_2 a_3}_{\m_2}(\vx_2) ... 
        A^{a_N a_{N+1}}_{\m_N}(\vx_N) \Bigr\} \overline{q}^{a_{N+1}}({\bf L})
\eea

   Before trying this approach in the full-blown field theory, it is
interesting to apply these ideas to a 
simple quantum-mechanical model with most
of the same qualitative features.  Let us take the
multi-gluon Hamiltonian to be
\beq
      H = \sum_{n=1}^{N-1} \Bigl|{\bf p}_n\Bigr| + 
          \sum_{n=2}^{N-1} V(\vx_n-\vx_{n-1}) + V_{qg}(\vx_1) 
          + V_{qg}({\bf L}-\vx_{N-1}) 
\eeq
  The {\bf Product Ansatz} for the (N-1)-gluon chain ground state is
\beq
\Psi(\vx_1,\vx_2,...,\vx_{N-1})=A\prod_{i=1}^N\psi({\bf u}_i)
\eeq
where 
\beq
     {\bf u}_i = \vx_{i} - \vx_{i-1}
\eeq
and
\beq
\psi({\bf u})=e^{-u^2/2r^2} .
\eeq
with the constraint
\beq
     \vx_0 \equiv {\bf 0} ~~,~~~ \vx_N = {\bf L}
\eeq
After some work, one finds that the string tension
(= energy/unit length) is
\beq
{{\cal E}\over L}={1\over rR}\sqrt{{8\over\pi}} + 
               {1\over R}\langle V(u) \rangle 
\eeq
In particular, taking for $V$ the instantaneous Coulomb potential
$V(u) = -C_F \a_s/|u|$, we have
\beq
{{\cal E}\over L}={1\over rR}\sqrt{{8\over\pi}}
-{C_F\alpha_s \over R^2}
\mbox{erf}\left({R\over r}\right) .
\eeq
where both $r$ and $N$ (no.\ of gluons), or equivalently
$ r$ and $R=L/N$, are taken as variational parameters.

  Our second trial wavefunction, the {\bf String Ansatz}, 
is adapted from discretized light-cone string theory \cite{thornlcft}:
\beq
  a_m^i \Psi(\vx_1,\vx_2,...,\vx_{N-1}) = 0 ~~~~
    \left\{    \begin{array}{l}
         m=1,2,..,N-1 \cr
         i=1,2,3 \end{array} \right.
\eeq
with 
\bea
\vec{x}_k& =&
{{\bf L}\over N}k+\sqrt{2\over NT_0}\sum_{m=1}^{N-1}
{1\over\sqrt{2\omega_m}}\left(\vec{a}_m+\vec{a}_m^\dagger\right)
\sin\left({m\pi\over N}k\right)
\non \\  
\vec{p}_k& =& -i\sqrt{2T_0\over N}\sum_{m=1}^{N-1}
\sqrt{\omega_m\over{2}}\left(\vec{a}_m-\vec{a}_m^\dagger\right)
\sin\left({m\pi\over N}k\right)
\eea
where $ \omega_m=2\sin{m\pi\over2N}$.
Define $ r \equiv 2/\sqrt{\pi T_0}$.
For a Coulomb potential, we then find
\beq
{{\cal E}\over L}={1\over rR}{8\over \pi^{3/2}}
-{C_F\alpha_s \over R^2}
\mbox{erf}\left({R\over r}\right) 
\eeq
Comparison with the product ansatz shows that the
string ansatz achieves a slightly lower energy.

  To estimate the string tension, we begin from
\beq
{{\cal E}\over L} = {1\over R}\left[ {8\over \pi^{3/2}}{1\over r} 
       + \langle V({\bf u}) \rangle \right] .
\label{EL}
\eeq
and make the approximation that
$\langle V({\bf u}) \rangle = V(s)$,
where
\beq
      {1\over s} \equiv \left\langle{1 \over |{\bf u}|}
                                      \right\rangle
           = {1\over R} \mbox{erf}\left({R\over r}\right)
\eeq
Minimizing wrt $R$ and $r$ results in two
conditions 
\bea
      {\E \over L} & =&  {\pa s \over \pa R} F(s) 
\non \\
  {8\over \pi^{3/2}}{1\over r^2} & =& 
 {\pa s \over \pa r} F(s) 
\eea
which together imply
\beq
     \s = {\pa s \over \pa R} \left({\pa s \over \pa r}\right)^{-1} 
                {8\over \pi^{3/2}}{1\over r^2} 
\eeq
Given $ F(s) \equiv dV/ds$, we could determine $ r,R$ and 
$ \s$.  Even without $ F(s)$, we can still make a
``ballpark'' estimate of $ \s$. 
First, the static force increases beyond $ L=0.6 r_0$, so a constituent
gluon must appear between the quarks at that point, to stabilize the
force.  This suggests that $ R=0.3 r_0.$  
Secondly, the running coupling $ \a(s)$ must be O(1), if the growth
in potential energy over-compensates the falloff in kinetic.  This
argues for $ s \approx 0.6 r_0$ and $ r \approx 0.63 r_0$.  With
these guesses, one finds
$\s = 1.26/r_0^2$,
which is to be compared with the experimental value
$\s_{exp} = (430~\mbox{Mev})^2 = 1.18/r_0^2$. 
We conclude that the value of the string tension \emph{may} 
work out in a genuine
field-theory/variational calculation, but of course
this remains to be seen.

\section{String Behavior}

   Using the string ansatz wavefunction, one finds for the transverse
coordinate of the $k$-th gluon in the chain
\bea
\bra{0}{\bf x}_{k\perp}^2\ket{0}& =& 
{D-2\over2NT_0}\sum_{m=1}^{N-1}
{\sin^2(m\pi k/ N)\over\sin(m\pi/2N)}
\non \\  \non \\ 
& \sim&  {D-2\over2\pi T_0}\ln N
\non \\ \non \\
& =&  {r^2(D-2)\over8}\ln{L\over R}
\eea
which demonstrates the expected logarithmic broadening of
the flux tube with quark separation.  Note that there is no need
to insert a high-frequency cutoff; this is taken care of by the
discreteness of the gluon chain. 

   The total energy of the gluon chain in the string ansatz can
also be computed, and the result minimized wrt the variational
parameters.  The result is 
\bea 
{\cal E}& =& 
{L\over\pi^{3/2}}\int d^3u\ e^{-{\bf u}^2}{\vec{R}\over R}\cdot\nabla V
\left(\vec{R}+2\vec{u}\sqrt{\Delta_1}\right)
\non \\ \non \\
   &&  \qquad -{\pi\over24L}\left({4R\over r\sqrt{\pi}}\right)
-\left[1+{\pi\over2}\right]{4\over r\pi^{3/2}}
\eea
(see ref.\ \cite{chain} for details).
The first term on the rhs is the linear potential, the second is a
L\"uscher-like term.  For the bosonic string, the factor multiplying $ \pi/24L$
in the L\"uscher term would be $ D-2=2$. 
With our estimate above of $ R\approx r/2$ this
coefficient is estimated to be $ 2/\sqrt\pi\approx1.13$.  At this stage,
however, numerical values should not be taken too seriously.

\section{Numerical Studies}

\begin{figure}[h!]
\centerline{\scalebox{0.50}{\includegraphics{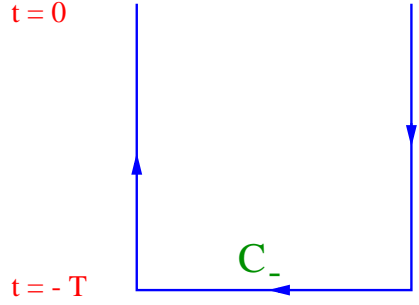}}}
\caption{Contour for generating the QCD string state.}
\label{cminus}
\end{figure}

   It is possible to investigate the constituent gluon composition
of the QCD string by numerical methods, and in fact a study along
these lines was carried out many years ago (second reference of ref.\
\cite{Jeff}).  The idea is the following:  Observe that the ground state
of the QCD flux tube, in a physical gauge, can be written as
\bea
   \lefteqn{\Psi_{string}[A_\m(x,t=0)] = } 
\non \\ 
     &&   \lim_{T \ra \infty} \Bigl( W(R,2T) Z\Bigr)^{-1/2}
        \int DA_\m(x,t<0) \d[F(A)] \Delta[A] 
\non \\
          && \qquad  \times P\exp[i\oint_{C_-} dx^\m A_\m]
        \exp\left[-\int_{-\infty}^0 dt ~L[A] \right]
\eea
where $C_-$ is the $R\times T$ open contour shown in Fig. \ref{cminus}.
The $R\times 2T$ Wilson loop factor $W(R,2T)$ ensures that $\Psi_{string}$
is normalized to unity.  

  If the gluon chain model is correct, then
$\Psi_{string}$ can also be expressed as a sum of n-gluon states
\bea
   \Psi_{string}[A] &=& \sum_n c_n \Psi_n[A]
\non \\
      \Psi_n[A] &=& N_n Q_n \Psi_0[A]
\eea
where $Q_n$ is an n-gluon operator, and
\beq
       N_n = \langle \oh \mbox{Tr}[Q^\dg_n Q_n] \rangle^{-1/2}
\eeq
is a normalization constant.  Assuming the $\Psi_n$ states are
orthogonal, the overlap of an n-gluon state with the true string
state is given, in the $T\ra \infty$ limit, by
\beq
  c_n = { \left\langle \mbox{Tr}[Q^\dg_n P\exp[i\oint_{C_-} dx^\m A_\m] 
      \right\rangle \over \sqrt{\left\langle \mbox{Tr}[Q^\dg_n Q_n] 
          \right\rangle W(R,2T)} }
\label{cn}
\eeq

   The second article of ref.\ \cite{Jeff}
used the following trial operators on the lattice 
(with $A_\m=(U_\m-U^\dg_\m)/2i$):  
\bea 
      Q_0^{ab} &=&  \d^{ab}
\non \\  
      Q^{ab}_1 &=&  \sum_{x_1<x<x_2} \overline{A}^{ab}_x(x)
\non \\  
      Q^{ab}_2 &=&  \sum_{x_1<x<x_2} \sum_{x<x'<x_2}
       \Bigl[ \overline{A}^{ac}_x(x) \overline{A}^{ca}_x(x') - \oh  
       \mbox{Tr}[\overline{A}_x(x) \overline{A}_x(x')] \Bigr]
\eea
\ni where 
\beq
   \overline{A}^{ab}_x(x) = \sum_{y,z} A^{ab}_x(x,y,z) e^{-\d r_\perp}
\eeq
is a ``smeared'' vector potential, $r_\perp$ is the transverse
distance from the flux tube axis, and $\d$ is a variational parameter.
The corresponding zero, one, and two-gluon overlaps ($c_0,~c_1,~c_2$)
were then computed from eq.\ \rf{cn} via lattice Monte Carlo.  It was
found that for small $R$, the zero-gluon overlap $c_0$ is the largest
of the three overlaps.  As $R$ increases, the $c_0$ overlap falls and
the $c_1$ term becomes dominant, and at the largest $R$ separations
that were used in the simulations we found the two-gluon overlap
$c_2$ becoming the largest of the three.  The sum of the three
overlaps accounts for about 90\% of the norm of $\Psi_{string}$, which
means that for the given range of $R$ the sum of zero, one and two
gluon states is a fairly good approximation to the true flux tube
state. For details, please see the cited reference.

   All of this is in good qualitative agreement with the gluon chain
picture, but the old results can surely be much improved.  It may be
useful to repeat the investigation of ref.\ \cite{Jeff} with modern
computers and better noise-reduction techniques.

\section{Conclusions}

   The gluon-chain model offers a simple and concise explanation of
many features of the confining force $-$ Casimir scaling at
large-$N_c$, center dependence, roughening, and the L\"uscher term $-$
which are problematic in many other approachs.  The model is
essentially a ``particle'' picture of string formation, and I regard
it as complementary to the ``field'' explanation of confinement in
terms of center vortices.  The next step will be to apply the field
theory/variational approach outlined above to obtain quantitative
estimates for the string tension and, perhaps, for the masses of the
low-lying glueballs.


\begin{thebibliography}{99}

%
\bibitem{thornweeparton}
C. B. Thorn, {Phys. Rev.} { D19} (1979) 639;
{ D20} (1979) 1435; { D20} (1979) 1934.
\bibitem{Jeff} J. Greensite, Nucl. Phys. { B249} (1985) 263; 
{ B315} (1989) 663.
\bibitem{GH} J. Greensite and M. Halpern, Nucl. Phys. B271 (1986) 379.
\bibitem{thornrpa}
C. B. Thorn, {Phys. Rev.} { D51} (1995) 647.
\bibitem{chain} J. Greensite and C. B. Thorn, J. High Energy Phys.
02 (2002) 014; hep-ph/0112326.
\bibitem{NS} S. Necco and R. Sommer, Phys. Lett. B523 (2001) 135;
hep-ph/0109093.
\bibitem{mrx} G. Grunberg, Phys. Rev. D40 (1989) 680.
\bibitem{Sommer} R. Sommer, Nucl. Phys. B411 (1994) 839.
\bibitem{Bachas} C. Bachas, Phys. Rev. D33 (1986) 2723.
\bibitem{Lat96} L. Del Debbio, M. Faber, J. Greensite, and S. Olejnik,
Nucl. Phys. Proc. Suppl. 53 (1997) 141, hep-lat/9607053.
\bibitem{j3} J. Ambj{\o}rn, J. Giedt, and J. Greensite,  J. High Energy
Phys. 02 (2000) 033, hep-lat/9907021.
\bibitem{thornlcft}
C. B. Thorn, {Nucl. Phys.} { B263} (1986) 493.

\end{thebibliography}
\end{document}